\documentclass[12pt,english]{article}
\usepackage[T1]{fontenc}
\usepackage[latin1]{inputenc}
\usepackage{geometry}
\usepackage{graphicx}
\usepackage{amssymb}

\makeatletter

\newcommand{\lyxaddress}[1]{
\par {\raggedright #1
\vspace{1.4em}
\noindent\par}
}

\usepackage{amsmath}
\usepackage{amssymb}

\makeatletter

\usepackage{babel}
\makeatother

\usepackage{babel}
\makeatother
\begin{document}

\title{Apparent density fluctuations in N-constant ensemble simulations}

\author{Aurélien Perera$^{1}$, Franjo Sokoli\'{c}$^{2}$ and Larisa Zorani\'{c}$^{1,*}$ }

\maketitle

\lyxaddress{$^{1}$ Laboratoire de Physique Théorique des Liquides (UMR CNRS
1600), Université Pierre et Marie Curie, 4 Place Jussieu, F75252,
Paris cedex 05, France.}

\lyxaddress{$^{2}$Laboratoire de Spectrochimie Infrarouge et Raman (UMR CNRS
8516), Centre d'Etudes et de Recherches Lasers et Applications, Université
des Sciences et Technologies de Lille, F59655 Villeneuve d'Ascq Cedex,
France.}

\begin{abstract}
In computer simulations performed in constant number of particles
ensembles, although the total number of particles N contained in the
simulation box does not fluctuate, hence giving a zero apparent compressibility,
there are still \emph{local} fluctuations in the number of particles.
It is shown herein that these apparent fluctuations produce a compressibility
that can be computed from the calculated radial distribution function,
and which matches to a great accuracy the compressibility of the fluid
for the open system.

This statement implies that the radial distribution function evaluated
in simulation of constant number of particles is identical to that
evaluated in the grand canonical ensemble, for the entire distance
range within half-box width. This is illustrated for the hard sphere
and Lennard-Jones fluids and for molecular models of water. The origin
of this apparent fluctuation is that the bulk of the remaining particles,
outside the range over which the distribution function is calculated,
act as a reservoir of particles for those within this range, thanks
to the periodic boundary conditions. The implications on the calculation
of the Kirkwood-Buff integrals are discussed.
\end{abstract}
\newpage

\section{Introduction}

The statistical theory of liquids is now mostly textbook knowledge.
Within such a theory one shows that the isothermal compressibility
of a liquid is related to the fluctuation in number of particles of
the system, through a fluctuation-dissipation type relation:

\begin{equation}
\frac{\chi_{T}}{\chi_{T}^{0}}=\left(\frac{\partial\rho}{\partial\beta P}\right)_{T}=\frac{<N^{2}>-<N>^{2}}{<N>}=1+\rho G\label{compr}\end{equation}
where $\chi_{T}^{0}=\beta/\rho$ is the ideal gas compressibility,
$\beta=1/k_{B}T$ is the Boltzmann factor and $\rho=N/V$ the number
density. The first equality in (\ref{compr}) is a thermodynamic definition,
while the two last equalities are derived in the grand canonical ensemble\cite{hansmac},
where the fluctuation in the number of particle is allowed. $G$ in
this equation is the integral of the radial distribution function
(RDF) $g(r)$\begin{equation}
G=4\pi\int_{0}^{\infty}dr\, r^{2}(g(r)-1)\label{G}\end{equation}

From Eq.(\ref{compr}) it is seen that the compressibility is related
to the fluctuation in number of particles through the second equality.
Therefore, we do not expect these relations to hold in any constant
N ensemble such as the micro-canonical (constant N,V,E), canonical
(constant N,V,T) or isobaric (constant N,P,T) ensembles. In such ensembles,
the last equality is expected to give $1+\rho G=0$. Many textbooks
show that this is indeed the case, when $G$ is evaluated in N-constant
ensemble. One such demonstration is provided in the next section of
this report. Since the RDF can be evaluated by simulations in different
ensembles, one expects the above constraint to be verified in N-constant
ensembles. We show here that this is not the case, and that, in fact,
the effective compressibility evaluated through Eq.(\ref{compr})
using the RDF, as well as the RDF itself, evaluated in the canonical
ensemble, for example, match the expected ones with great accuracy.
This contradiction is lifted by considering contributions from the
periodic boundary conditions, as discussed later. In fact, this result
should not surprise us, since it is well known that the chemical potential
can be evaluated from simulations in N-constant ensemble\cite{widom}.
The procedure consist in evaluating the insertion free energy of an
additional particle, which implicitly supports the existence of local
density fluctuations. It is these local fluctuations that give rise
to the apparent global density fluctuation, in the very absence of
any such macroscopic feature.

Such behaviour is observed equally in mixtures, where, in addition
to density fluctuation, concentration fluctuation also play an important
role. Hence, it gives some additional support for the evaluation,
through computer simulations, of the so-called Kirkwood-Buff integrals
(KBIs)\cite{KB}, that have attracted recent interest in the modeling
of the force field of aqueous mixtures\cite{smith,smith2}. Indeed,
such mixtures tend to show appreciable micro-segregation, that can
be detected to some extent through the comparison with the experimental
KBIs\cite{smith,ours,ours2}. Since these quantities are simply the
integrals of the various site-site radial distribution functions,
the evaluation of the latter by computer simulations has an undeniable
interest. However, the theory behind the KBIs is strictly a grand
canonical approach\cite{KB}. Then, one is interested to know what
are the limitations of the RDF computed in constant-N ensemble simulations

It is worthwhile mentioning that, the problematic of comparing the
RDF evaluated in different ensembles, have been addressed in the past
by few authors\cite{petitt,benNaim}. To our knowledge, however, the
fact that the isothermal compressibility and the RDF match that evaluated
from grand canonical ensemble has not been addressed in present terms.

The paper is organized as follows; the next section presents the theoretical
material that is needed to clarify the issue raised here . The results
section contains an illustration on the hard sphere, Lennard-Jones
and some water models. Finally the discussions and conclusions are
given in section 4.

\section{Theoretical details}

As stated in the introduction, there is a connection between the fluctuation
of the number of particles N and the isothermal compressibility $\chi_{T}$,
which is found in many textbooks of the liquid state theory\cite{hansmac}.
We recall here briefly the main steps for completeness sake.

The RDF is a key quantity in the analysis of liquids since it provides
a direct information about the microstructure of liquids. This function
is related to the second member of the whole hierarchy of density
correlation functions, that can be constructed as statistical moments
from one single function, the microscopic density in a N-particles
system\begin{equation}
\rho(1)=\sum_{i=0}^{N}\delta(i-1)\label{rho1}\end{equation}
 with $\delta(i)$ being the Dirac delta function and with the convention
that $1=(\vec{r}_{1},\underline{\Omega}_{1})$ represents the position
and the orientation of molecule labeled $1$. This definition is equivalent
to a snapshot of the instantaneous position and orientation of the
particles, and thus represents \emph{one microstate} of the system.
It is interesting to note that this quantity is the basic observable
that can be perused by computer simulations. In practice, however,
we are essentially interested in averages of this quantity and its
various correlations. These averages can be performed in different
ensembles. For a given ensemble, one can define, for example, the
one-body function $\rho^{(1)}(1)=<\rho(1)>$ which for the translationally
and rotationally invariant systems (homogeneous and isotropic) that
we consider in this report is simply\begin{equation}
\rho^{(1)}(1)=<\rho(1)>=\bar{\rho}=\rho/\omega\label{rho}\end{equation}
 that is the number density $\rho=N/V$ divided by the solid angle
$\omega$ (where $\omega=4\pi$ or $8\pi^{2}$ depending on the symmetry
of the molecule).

Going one step further, one can equally define the second moment,
namely the two particles density as\begin{equation}
\rho^{(2)}(1,2)=<\rho(1)\rho(2)-\delta(1,2)\rho(1)>=<\sum_{i\neq j}\delta(i-1)\delta(j-2)>\label{rho2}\end{equation}
 The pair correlation function $g(1,2)$ is then defined by\begin{equation}
\rho^{(2)}(1,2)=\bar{\rho}^{2}g(1,2)\label{g12}\end{equation}
 This definition implies that when the particles are uncorrelated,
that is for example when they are infinitely far apart, then one has
exactly $\rho^{(2)}(1,2)=\bar{\rho}^{2}$, which means:\begin{equation}
\lim_{r\rightarrow\infty}g(1,2)=1\label{limg12}\end{equation}
 It is clear that this relationship is trivially valid for an infinite
system, but must be revised for a finite system with N particles in
a volume V. This is, for example, the case of the micro-canonical
(constant NVE), canonical (constant NVT) or isobaric (constant NPT)
statistical ensembles, in which most simulations are performed. In
the following discussion, we take as example the canonical ensemble,
but the reasoning equally holds for the other two ensembles.

For the translationally and rotationally invariant systems considered
in this report, equation (\ref{g12}) together with (\ref{rho1})
leads to\begin{equation}
g(1,2)=\frac{N(N-1)}{\bar{\rho}^{2}Z_{N}}\int d3...dN\exp(-\beta\mathfrak{V}(N))\label{gcano}\end{equation}
 where $Z_{N}=\int d1...dN\exp(-\beta\mathfrak{V}(N))$ is the canonical
ensemble partition function, with $\mathfrak{V}(N)$ being the total
interaction energy between the N particles. The RDF $g(r)$ is then
defined as the angle average of $g(1,2)$ as:

\begin{equation}
g(r)=\frac{1}{\omega^{2}}\int d\underline{\Omega}_{1}d\underline{\Omega}_{2}g(1,2)\label{gr}\end{equation}
 It is perhaps worthwhile noting here that the definition (\ref{gcano})
contains the factor $N(N-1)/N^{2}$, where the numerator arises from
the number of pairs in an ensemble of $N$ particles, while the denominator
is coming from the factor $\bar{\rho}^{2}$ in (\ref{g12}). It is
this factor that is responsible for the asymptotic limit of the RDF
in finite systems.

Let us examine the asymptotic behaviour from Eqs.(\ref{gcano},\ref{gr})
when $r\rightarrow\infty$. From Eq.(\ref{gcano}), when $r_{12}\rightarrow\infty$
, that is when the two particles are far apart, one can neglect the
interaction between them in $\mathfrak{V}(N)$, hence one gets directly
from Eqs.(\ref{gcano},\ref{gr}), for the canonical ensemble\begin{equation}
\lim_{r\rightarrow\infty}g(r)=1-1/N\label{limgr}\end{equation}
 This relation hold equally in the micro-canonical ensemble. In the
isobaric, constant NPT ensemble, the corresponding limit is $g(r)\rightarrow(1-1/N)\lambda(V)$,
where $\lambda(V)$ is a formal function whose form is given in the
appendix. This size dependence is expected to be observed in the canonical
ensemble, and one can see from Eq.(\ref{compr}) that the integral
$G$ should trivially satisfy\begin{equation}
1+\rho G=0\label{closed}\end{equation}
 since N does not fluctuate.

The RDF can also be directly obtained from computer simulations from
Eq.(\ref{rho2}). This equation tells us that we only need to compute
the histogram $H(r,\Delta r)$ which counts the pairs of particles
at a distance between $r$ and $r+\Delta r$. The resulting expression
is:\begin{equation}
g(r)=\frac{H(r,\Delta r)}{N^{2}\Delta V(r,\Delta r)}\label{gsimu}\end{equation}
 where $\Delta V(r,\Delta r)=(4\pi r^{2}\Delta r)/V_{0}$ is the volume
of the spherical element considered, reduced by the total volume $V_{0}$
of the simulation cell. The above expression is the equivalent of
the second equality in Eq.(\ref{rho2}). In practice, this calculation
is performed and averaged over several configurations or microstates.
It is very important to note that Eq.(\ref{gsimu}) is valid for any
statistical ensemble. It is the nature of fluctuations of a particular
ensemble that will affect the form the histogram $H(r,\Delta r)$
and determine the corresponding asymptotic behaviour of $g(r)$. Hence,
it is not appropriate to introduce additional factors into this expression,
in order to enforce expected behaviour. This point is worth mentioning,
since incorrect additional factors are often found in some algorithms.

An important remark must be made at this point. The computer simulations
of the various N constant ensembles are not strictly concerned by
arguments about finite or closed systems, since the periodic boundary
conditions allow to mimic an infinite system. This is, of course,
size dependent. It turns out that this fact permits some kind of pseudo-fluctuations
in the number of particles within specific subvolumes, that we consider
to be local fluctuations, while the total number of particles in the
simulation cell does not fluctuate. In order to grasp what this feature
could induce, let us imagine a very large system, that is constrained
to be finite (a fluid in a huge box). Then, we consider computing
the g(r) in a very small sub-system inside this large system. Clearly,
the outer part of the system acts as a particle reservoir, and we
naturally expect the resulting g(r) to contain the fluctuation of
particles correctly described, that is the correct asymptotic limit,
\emph{i.e.} unity. Since the g(r) of small systems are computed within
a sub cell of the system box, usually in a spherical shell with radius
about L/2, where L is the simulation box size, we expect some influence
on the g(r) of the fluctuations of the number of particles of this
subsystem. This is what we examine in the next section. It should
be clear, at this point, that our considerations about apparent fluctuations
hold only for subvolumes smaller than that of the largest sphere inscribed
within the cubical box, and that fluctuations beyond this range will
inevitably show that the total number of particles is fixed. 

These definitions can be equally extended to mixtures. In particular,
the Kirkwood-Buff integrals are defined between species labeled $\alpha$
and $\beta$ as:

\begin{equation}
G_{\alpha\beta}=4\pi\int_{0}^{\infty}dr\, r^{2}(g_{\alpha\beta}(r)-1)\label{kbi}\end{equation}
 where $g_{\alpha\beta}(r)$ is the corresponding RDF.

\section{Results}

Nothing, in the general formalism outlined above, prepare for the
actual results we have observed in our simulations. We have evaluated
the RDF and the resulting compressibility (through (\ref{compr}))
for the hard sphere fluid and two water models. Constant NVT Monte
Carlo (MC) simulations have been performed for the hard sphere and
Lennard-Jones fluids, for various system sizes (N=500, 2048, 4000).
These two models are the archetypal models for simple liquids, and
thus serve as a perfect basis to test the claims. Molecular Dynamics
simulations in the constant NPT ensemble have been performed for the
SPC/E\cite{spce} and TIP5P\cite{tip5p} water models, with N=864,
2048 and N=10976. The calculations show in an unbiased manner that
the RDF have the correct asymptotic limit, \emph{i.e.} unity.

\subsection{The hard sphere fluid}

One component hard sphere (HS) fluid has been equilibrated using a
constant NVT Monte Carlo code, for various system sizes, and for the
fixed reduced number density $\rho*=(N/V)\sigma^{3}=0.8$, where $\sigma$
is the hard sphere diameter. This density can be considered as dense
enough for a liquid phase. The RDF for systems with N=500, 2048 and
10976 have been computed with statistics performed over 100 million
particle moves. This exceeds by far the usually encountered numbers,
but it is necessary in order to have accurate evaluation of the g(r)
at large distances.

Fig.1 shows the RDFs in the entire range, and the inset shows the
integrand $y(r)=(g(r)-1)r^{2}$. The expected asymptotes $-r^{2}/N$
are also plotted, each for the range over which the RDFs have been
evaluated. The inset indicates clearly that $y(r)$ oscillates around
zero, rather than the expected asymptotes. The RDF calculated from
the Percus-Yevick (PY) theory is equally shown. Since this RDF is,
by definition, calculated in the grand canonical ensemble, it serves
as a test of the asymptotic limit, despite being an approximate theory.
One notices in particular the very good agreement at large distances
between the oscillatory structure from the PY theory and the simulations
with N=10976, which is clearly seen in the inset.

\subsection{The Lennard-Jones fluid}

The one component Lennard-Jones (LJ) fluid has been also studied by
an NVT constant MC algorithm, for reduced density $\rho^{*}=0.9$
and reduced temperature $T^{*}=k_{B}T/\epsilon=3.0$ (where $\epsilon$
is the energy depth of the LJ interaction). This choice corresponds
to a point in the dense and high temperature state of the LJ fluid.
Systems sizes of N=500, 2048 and 10976 have been studied, with statistics
similar to that of the hard sphere fluid.

Fig.2 is the analogous of Fig.1 but for the LJ fluid. Once again,
it is observed that the RDF tends to unity, despite the fact that
we are in an constant N ensemble. The PY results are also plotted
and serve as a reference, despite the approximate nature of this theory.
In the high temperature region, this theory is expected to be more
appropriate than the hypernetted chain (HNC) theory, since in the
high temperature regime the correlations are expected to be dominated
by excluded volume effects, and be closer to the HS regime\cite{hansmac}.
It is observed that the long range oscillatory structures, between
the simulations and the theory, are again in very good agreement.
The large oscillations for the N=10976 system are due to the high
temperature, and are difficult to reduce, despite statistics collected
over 100 million moves. Again it is observed that none of the data
follow the expected 1/N asymptotes that are plotted in the inset.

Fig.3 shows the running reduced compressibility $\chi_{T}^{*}=\chi_{T}/\chi_{T}^{0}$
for the same system. The numerical value obtained from the PY theory
is $\chi_{T}^{*}=0.05254$, shown as an horizontal line, and serves
as a comparison point for the unknown exact compressibility for this
state point. The agreement between the asymptotic limit evaluated
through Eqs.(\ref{compr},\ref{G}) indicates that the pseudo macroscopic
density fluctuation in this system where neither N nor V are changing,
tend to a non zero value, which is quite nicely seen for the N=10976
system. This value of the compressibility, estimated to be about $0.035$,
is below both that of the approximate PY and HNC(0.07564) theories,
which tends to overestimate the correct compressibility. The virial
pressures (compressibility factor $Z=\beta P/\rho$) obtained from
the HNC and PY theories are 6.398 and 4.90, respectively, while that
obtained from the simulations is 5.140. The reduced energies per particle
are -1.167 (HNC), -1.453(PY) and -1.395 (MC). The simulated values
are seen to be in between those of the two theories. This reflects
the trend generally observed that the two theories bracket the exact
results. 

As a side remark, it is rather worthy of note that the convergence
of the value of the compressiblity should need systems sizes as large
as N=10976 for systems as simple as the HS or LJ fluids. Properties
such as the pressure or the internal energy, converge faster, that
is within a smaller range of distance. The typical example of this
is the pressure for hard spheres that requires only the contact value
of the g(r). The principal reason for these properties to converge
faster is that the integrand is weighted by the interaction (or the
derivative), which screens the long range oscillatory behaviour of
the RDF.

\subsection{Water models}

Molecular Dynamics simulations in constant NPT ensemble, using the
DL-POLY2 code\cite{dlpoly}, have been performed for two water models,
namely the SPC/E and TIP5P models, under ambient conditions. The oxygen-oxygen
RDF $g_{OO}(r)$have been computed for several system sizes, which
is N=864, 2048 and 10976 for the SPC/E model, and N=2048 for the TIP5P
model. Averages have been performed over several hundreds of thousand
steps.

Fig.4 shows the RDF, and the inset shows the integrand $(g_{OO}(r)-1)r^{2}$.
Again, the asymptotic limit is near perfectly unity, in contradiction
with the fact that this calculation is done in the N,P,T constant
ensemble, where the strict application of the arguments developed
in Section 2, give the following expectation, $\lim_{r\rightarrow\infty}g(r)=(1-1/N)\lambda(V)$
(see the appendix).

Fig.5 shows the experimental and calculated KB integrals from simulations.
The experimental compressibility at room temperature is $0.4533GPa^{-1}$\cite{compress}.
From this value we estimate the experimental G to be $G=-16.9cm^{3}/mol$,
which is shown as an horizontal line. The agreement is very good,
indicating again that the true and apparent compressibilities are
in good correspondence. One can note the fact that both water models
reproduce well the experimental compressibility, which is an indication
of their accuracy with that respect (TIP4P is equally in the same
accuracy region, although not shown here).

\section{Discussion and conclusion}

The results shown in the preceding section demonstrate numerically
that the apparent density fluctuations in N constant ensemble simulations
match near perfectly those of a corresponding system where N would
be allowed to fluctuate. In addition, the RDF evaluated in such ensemble
tend to the correct asymptotic behaviour, \emph{i.e.} unity. These
findings are purely empirical, and nothing in the theoretical considerations
of Section 2 allows us to anticipate or justify these findings.

Examining the conditions of the realization of the various N-constant
ensemble in computer simulations, we notice that these ensembles are
not {}``closed'', since they are made infinite through the periodic
boundary conditions. In fact, a \emph{finite} constant N system cannot
be {}``closed'' in the sense of having a fixed boundary, since this
boundary will always influence the distribution of the particles by
making the system inhomogeneous, and thus rendering useless the formalism
exposed in the section 2. In this, the term {}``closed'' used in
Ref.\cite{benNaim}, and various subsequent reports by other authors,
is totally inappropriate. The absence of boundaries permits an effective
fluctuation of the number of particles inside any subvolume, and the
numerical evidence here simply indicates that this fluctuation is
enough to induce an apparent macroscopic fluctuation through the RDF,
that is near exactly that of the open system. The fact that a small
system of few hundreds to few thousands particles can reproduce a
macroscopic quantity is not new; many macroscopic thermodynamic quantities
are evaluated within such small microscopic subsystems that are the
simulation cells. In fact, as stated in the introduction, the very
fact that one can use an N-constant cell to compute the chemical potential,
indicates that the fluctuations in the system are the appropriate
response to local variations in pressure and volume. 

One of the interesting conclusions of this study is the perspective
of its application to mixtures, and particularly to aqueous mixtures.
Indeed, in such systems, concentration fluctuations are very important,
due to specific self-clustering tendencies induced by the ability
of water molecules to link together through hydrogen bonds. One of
the questions that have aroused in recent works \cite{ours,ours2}
is the ability of the constant N simulations to reproduce properly
the correct long range behaviour of the various RDFs, and in particular
that between water molecules. The present study gives some confidence
in the constant N simulation studies of such systems, once care is
taken to consider sizes large enough to support local immiscibility
without leading to macroscopic phase separation.

\section*{Appendix}

The computation of the formal limit of g(r) in the NPT ensemble follows
a logic similar to that exposed in Section 2. We start from the equivalent
of (8) which is now\begin{equation}
g(1,2)=\frac{N(N-1)}{\tilde{\rho}^{2}Q_{N}}\int_{0}^{\infty}dV\exp(-\beta PV)\int d3...dN\exp(-\beta\mathfrak{V}(N))\label{a1}\end{equation}
where $Q_{N}=\int_{0}^{\infty}dV\exp(-\beta PV)\int d1...dN\exp(-\beta\mathfrak{V}(N))$
is the partition function of the (N,P,T) constant isobaric ensemble\cite{hansmac}.
The number density $\tilde{\rho}$ is now defined as $\tilde{\rho}=N/\tilde{V}$,
where $\tilde{V}$ is the effective average volume of the system under
constant pressure $P$. When $r_{12}\rightarrow\infty$, that is when
the interaction between particles 1 and 2 can be neglected, Eq.(\ref{a1})
reduces to\begin{equation}
g(1,2)\rightarrow\frac{N(N-1)}{\tilde{\rho}^{2}}\frac{1}{<V^{2}>^{'}}\label{a2}\end{equation}
where \begin{equation}
<V^{2}>^{'}=\frac{\int_{0}^{\infty}dV\; V^{2}\exp(-\beta PV)f(V)}{\int_{0}^{\infty}dV\exp(-\beta PV)f(V)}\label{a3}\end{equation}
with $f(V)=\int d3...dN\exp(-\beta\mathfrak{V}'(N))$. The $^{'}$
indicates that the volume average is taken for (N-3) particles only.
Eq.(\ref{a2}) can be rewriten as \begin{equation}
g(1,2)\rightarrow(1-\frac{1}{N})\frac{\tilde{V}^{2}}{<V^{2}>^{'}}\label{a4}\end{equation}
The second fraction involving the volumes can be evaluated easily
in the ideal gas limit, but is not so for the case the fully interacting
system. Nevertheless, one sees that the asymptotic limit of the isotropic
part of g(1,2) is not trivially unity, as would be expected in an
open system.

\newpage

\section*{Figure captions}

\begin{enumerate}
\item \textbf{Fig.1}. (\emph{color online}). RDF of the hard sphere fluid
at $\rho^{*}=0.8$. MC simulation results for N=500 (green), N=2048
(blue) and N=10976 (red). PY result in black. Inset: plots of $y(r)=(g(r)-1)r^{2}$
with same color conventions. 
\item \textbf{Fig.2}. (\emph{color online}). RDF of the Lennard-Jones fluid
at $\rho^{*}=0.8$ and $T^{*}=0.9$. MC simulation results for N=500
(green), N=2048 (blue) and N=10976 (red). HNC result in black. Inset:
plots of $y(r)=(g(r)-1)r^{2}$ with same color conventions. 
\item \textbf{Fig.3}. (\emph{color online}). Running (reduced) compressibility
integral (see text) for the system shown in Fig.2, with same color
convention. The horizontal line is the corresponding HNC compressibility.
\item \textbf{Fig.4}. (\emph{color online}). Water oxygen-oxygen RDF from
MD simulations for pure water at ambient conditions. SPC/E with N=864
(magenta); SPC/E with N=2048 (blue); SPC/E with N=10976 (black); TIP5P
with N=2048 (cyan). The inset shows the corresponding $y(r)$.
\item \textbf{Fig.5}. (\emph{color online}). The running KB integral for
the systems shown in Fig.4, with same conventions. The horizontal
line is the experimental result (see text).
\end{enumerate}
\newpage

\includegraphics[scale=0.8]{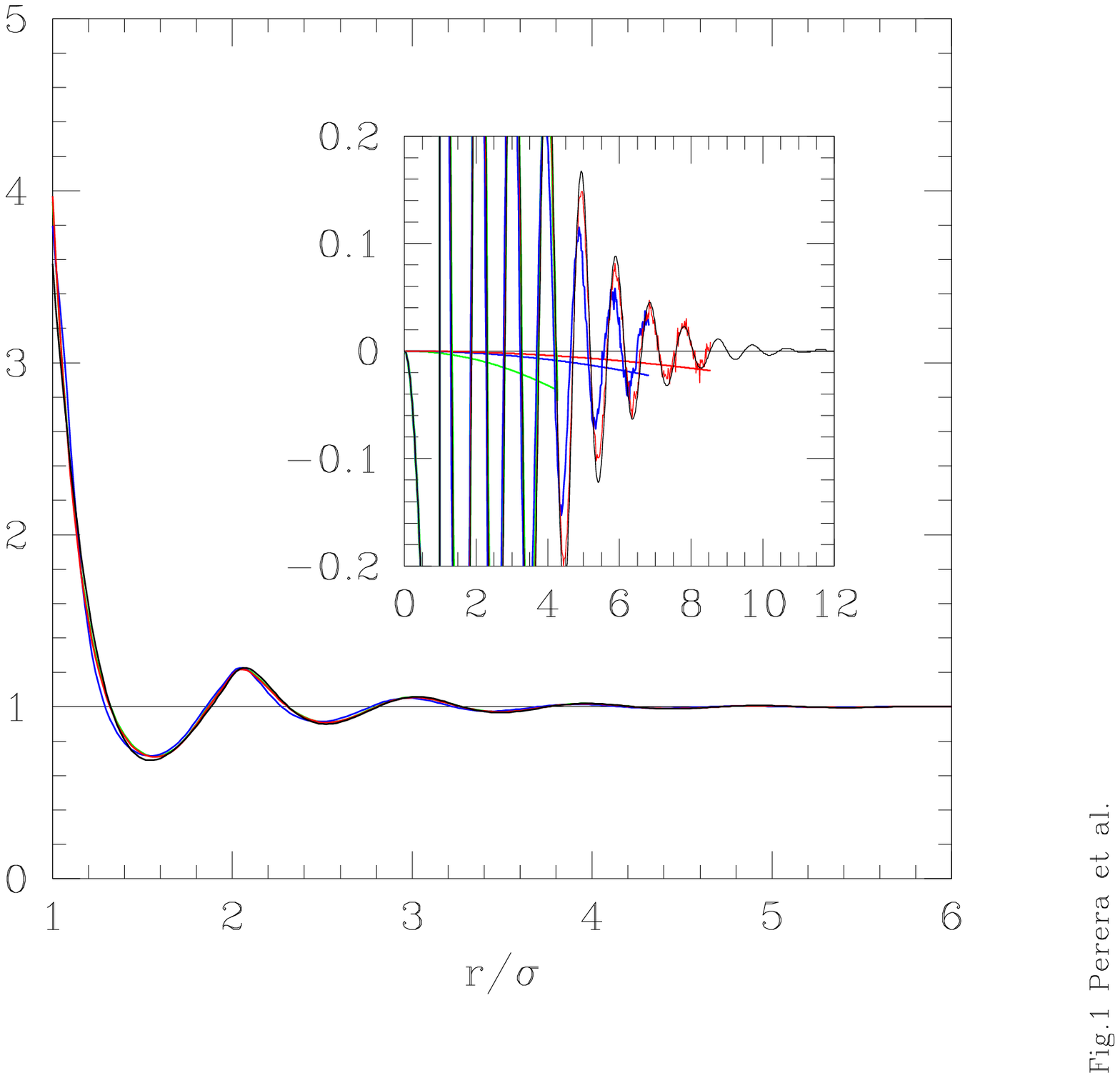}

\newpage

\includegraphics[scale=0.8]{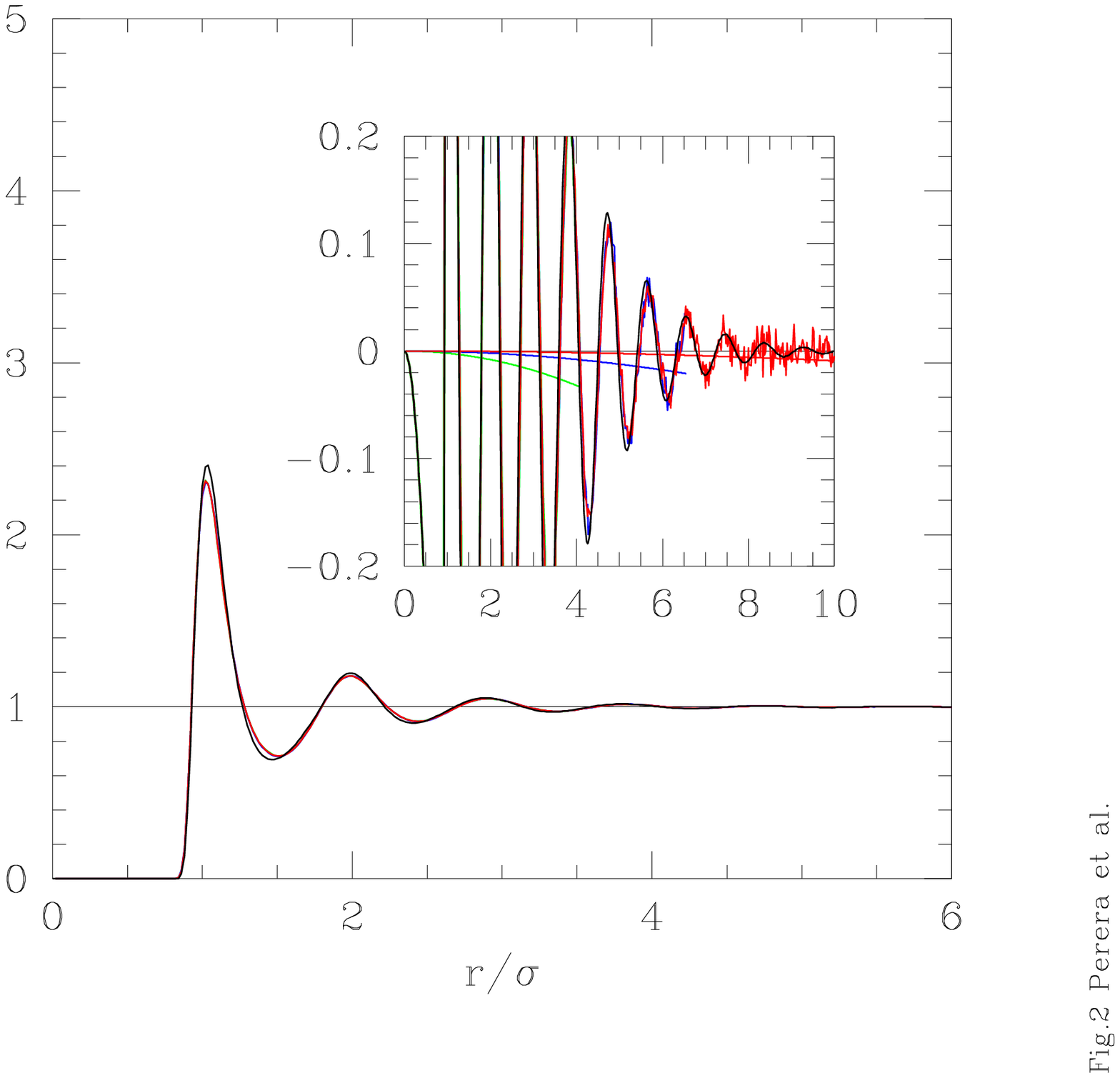}

\newpage

\includegraphics[scale=0.8]{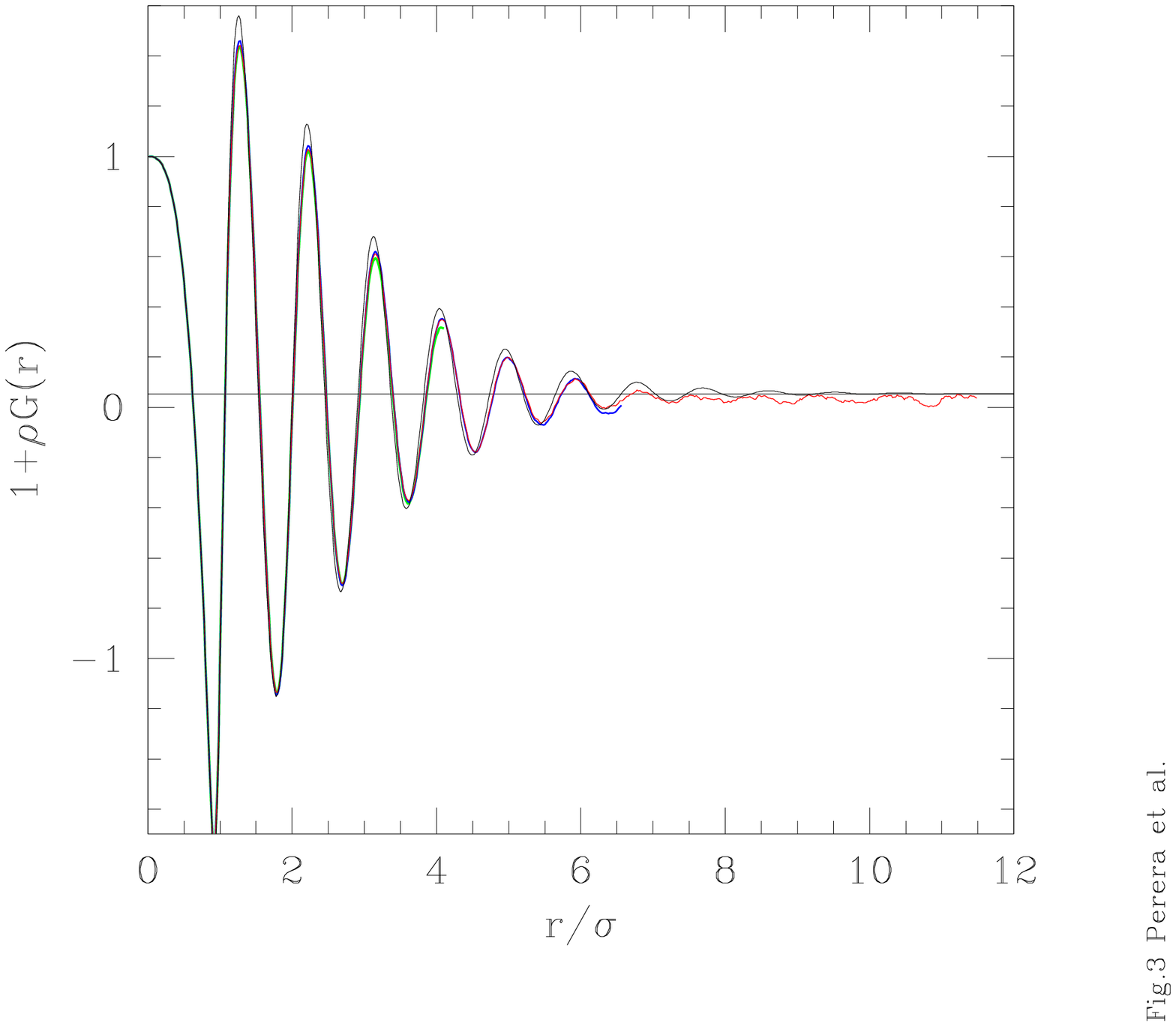}

\newpage

\includegraphics[scale=0.8]{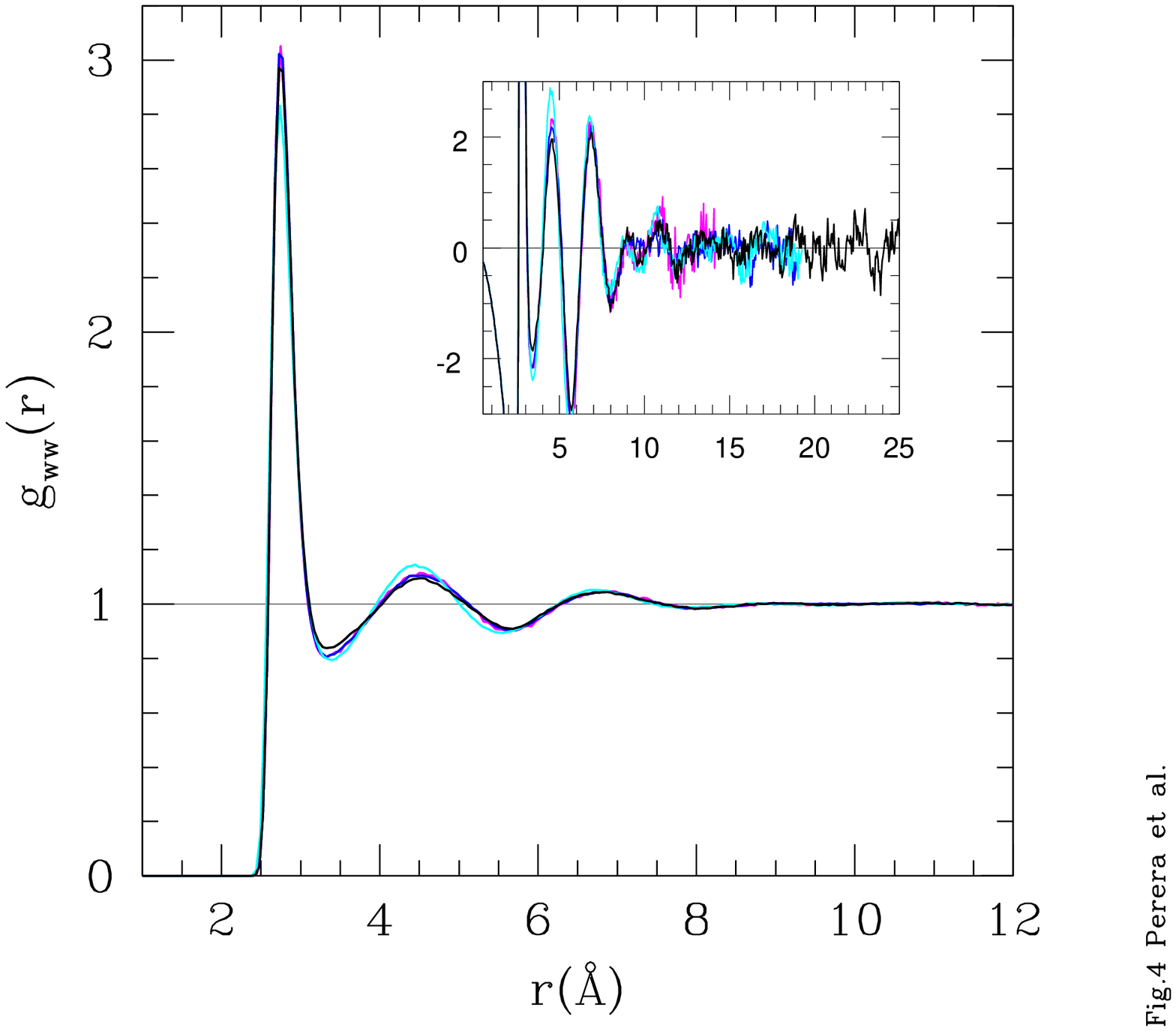}

\newpage

\includegraphics[scale=0.8]{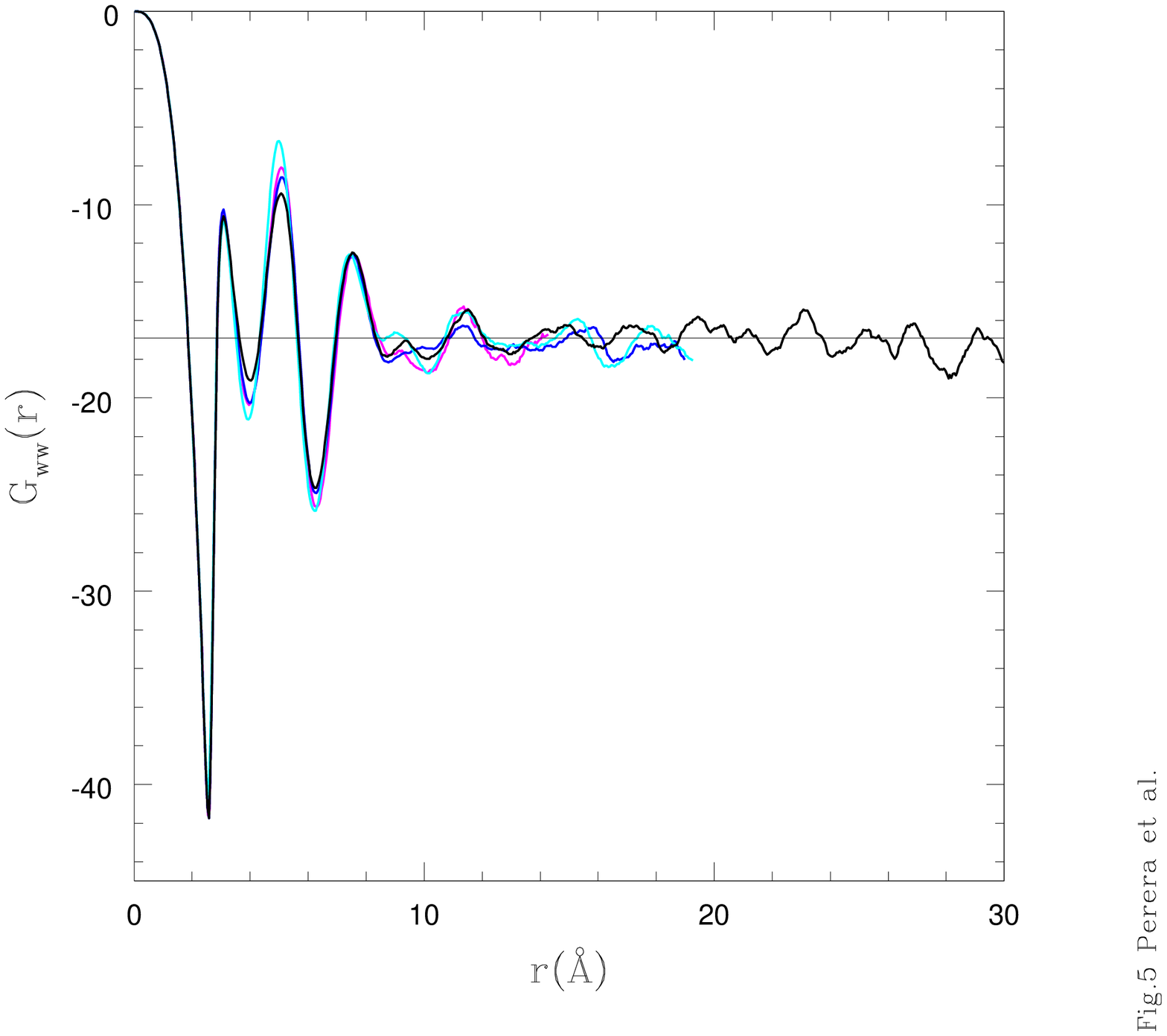}
\end{document}